%% file: main.tex
\documentclass[final,5p,times,twocolumn,numbers, compress]{elsarticle}

\usepackage{amssymb}

\journal{Nuclear Instruments and Methods in Physics Research Section A}

\include{header}

\include{mycommands}

\begin{document}

\begin{frontmatter}



\title{Bias-Free Estimation of Signals on Top of Unknown Backgrounds}


\author[MPP,TUM]{Johannes Diehl}
\ead{diehl@mpp.mpg.de}
\author[TUM,ODSL,RU]{Jakob Knollmüller}
\ead{jakob.knollmueller@ru.nl}
\author[MPP,ODSL]{Oliver Schulz}
\ead{oschulz@mpp.mpg.de}

\affiliation[MPP]{organization={Max-Planck-Institut for Physics},
            addressline={Foehringer Ring 6}, 
            city={Munich},
            postcode={80805}, 
            country={Germany}}
\affiliation[TUM]{organization={Technical University of Munich, TUM School of Natural Sciences},
            addressline={Boltzmannstr. 2}, 
            city={Garching},
            postcode={85748}, 
            country={Germany}}
\affiliation[ODSL]{organization={ORIGINS Data Science Lab, Excellence Cluster ORIGINS},
            addressline={Boltzmannstr. 2}, 
            city={Garching},
            postcode={85748}, 
            country={Germany}}

\affiliation[RU]{organization={Department of Astrophysics, Radboud University},
            addressline={Heyendaalseweg 135}, 
            city={Nijmegen},
            postcode={6525 AJ}, 
            country={Netherlands}}

%
%

\begin{abstract}
    We present a method for obtaining unbiased signal estimates in the presence of a significant unknown background, eliminating the need for a parametric model for the background itself. Our approach is based on a minimal set of conditions for observation and background estimators, which are typically satisfied in practical scenarios. To showcase the effectiveness of our method, we apply it to simulated data from the planned dielectric axion haloscope MADMAX.
\end{abstract}




\end{frontmatter}




\section{Introduction}
\label{sec:Introduction}
\vs{-2mm}
Fitting a small-amplitude signal in the presence of a large-amplitude background is both a common and often challenging problem. If one has a valid parametric model for both signal and background, and the response of the experimental apparatus can be accurately modelled as well, then a forward-modelling approach can be employed: With signal parameters $\theta$ and background parameters $\phi$ we can usually construct a tractable and parameterised probability distribution $p^{\obs}_ {\theta, \phi}(X)$ that models the probability of observing a specific realisation of $X$. The combination of such a distribution with some actual observed data results in a likelihood function, and so all the common tools of frequentist and (with priors on signal and noise) Bayesian statistics can be brought to bear. If, however, a parametric model is only available for the signal, but not for the background, the situation is less straightforward.

In some cases it is possible to describe the background fairly accurately with a flexible empirical function. The resulting fit of signal and background is then bias-free. This approach does, however, add a potentially large number of nuisance parameters, depending on the complexity of the background and the dimensionality of the data. As a result, the numerical cost of the fit can increase substantially and achieving fit convergence can be a challenge. Also, it's often not easy to ensure that such an empirical function will indeed cover all possible background shapes.

Alternatively, one can use a parameter-free background filter, equivalent to subtracting a parameter-free background estimate from the observation. Unfortunately, such an estimator is typically affected by the presence of a signal, resp.\ such a filter does alter the signal to some degree. As a result, a subsequent signal estimate will be biased unless additional measures are taken to correct for this.

This issue does, for example, arise in the context of axion haloscope experiments. These aim to detect a small, localised axion signal on top of a dominating radio-frequency background. This background is determined by the system response and characteristics of the radio-frequency receiver chain of the haloscope experiment. As the wavelengths of interest are comparable to the size of the system, it is exceedingly difficult to model this background ab-initio.

Often Savitzky-Golay or similar filters are used to non-parametrically subtract the background while simultaneously retaining potential signals \cite{Palken:2020wgs, Beaujean:2017eyq}. But such a background filter also affects the signal, changing its amplitude and shape. This leads to a bias if signal parameters are inferred directly from the filter output. A common method to correct for this bias uses simulated pseudo-experiments \cite{ATLAS:2020ocz} to estimate the filter-induced distortion of the signal on an ensemble-level. In an axion context this has so far only been done to correct the bias on the total signal power \cite{ADMX:2020hay, Brubaker:2017rna}, but not regarding axion parameters coupled to the signal shape. Recent efforts to improve upon the standard analysis pipeline exist \cite{Palken:2020wgs, GalloRosso:2022mhx}, but do not tackle this effect.

In this work we demonstrate an approach that does use parameter-free background filters but still yields inherently unbiased signal estimates. We explain the general principle of our approach in Sec.~\ref{sec:approach}. In Sec.~\ref{sec:application} we demonstrate the approach by applying it to a physics example in context of the dielectric axion haloscope MADMAX \cite{PhysRevLett.118.091801, Millar:2016cjp, MADMAX:2019pub, Beurthey:2020yuq} using Savitzky-Golay filters as background estimators. We conclude in Sec.~\ref{sec:Summary_and_Outlook}.
\vspace{-2mm}
\section{General Approach}
\label{sec:approach}
\vs{-2mm}

While the approach described here is fairly generic, we do require a few condition to be fulfilled in order to gain unbiased signal estimates:
\begin{itemize}
\item The expected value $E(X)$ of the experimental observation $X$ can be written as linear combination of a signal $S$ and a background $B$:
\begin{equation}
E(X) = S + B.
\end{equation}
\item The response of the experiment, i.e., the measure of probability $p^{\obs}(X)$ of observing an outcome $X$, can be modelled with sufficient accuracy by a tractable probability distribution that is parameterised by its expectation $E(X)$:
\begin{equation}
p^{\obs}_{E(X)}(X) = p^{\obs}_{S+B}(X).
\end{equation}
If so, then we can also split the observation X into signal plus background $S +B$ and noise $N$ and write $p^{\obs}$ (without loss of generality) as
\begin{equation}
p^{\obs}_{S+B}(X) = p^{\noise}_{S+B}(N),
\end{equation}
with
\begin{equation}
\label{eq:signal_decomposition}
X = E(X) + N = S + B + N
\end{equation}
and $E(N) = 0$.
\item We have a parameter-free, unbiased and effective background estimator $f_{\bg}$, i.e. we require it to estimate the background faithfully (relative to possible signals amplitudes and length-scales):
\begin{equation}
\label{eq:bg_est_approx}
f_{\bg}(B) \approx B \quad \mathrm{e.g.} \quad |B - f_{\bg}(B)| \ll |\mathrm{ampl}(S)|
\end{equation}
and we require it to suppress noise effectively (relative to the noise level):
\begin{equation}
\label{eq:bg_est_noiseremoval}
f_{\bg}(N) \approx 0 \quad \mathrm{e.g.} \quad |f_{\bg}(N)| \ll \sqrt{\mathrm{var}(N)}
\end{equation}

Less formally, $B - f_{\bg}(B)$ must not be allowed to mimic a signal and $X - f_{\bg}(X)$ must have approximately the same noise as $X$.
\item The background estimator is linear: 
\begin{equation}
\label{eq:bg_est_linear}
f_{\bg}(S + B + N) = f_{\bg}(S) + f_{\bg}(B + N)
\end{equation}
\item The background estimator does not reconstruct non-zero signals perfectly:
\begin{equation}
\label{eq:keep_some_signal}
S \neq 0 \implies f_{\bg}(S) \neq S,
\end{equation}
so that background removal does not completely eliminate signals. But crucially, we do \emph{not} require the signal to be invariant under background removal, that means we allow for $S - f_{\bg}(S) \neq S$.
\item The possible shapes of the signal $S$ are known and so $S$ can be parameterised by signal parameters $\theta$ and expressed as a tractable $S_{\theta}$.
\end{itemize}
The first two requirements here are usually satisfied in signal-plus-background inference scenarios anyway. The central requirements are the existence of an unbiased, effective and linear background estimator and a tractable parameterisation of the signal shape.

Note that the domain of $S$ and $B$ may be different than the domain of $X$. If, e.g., $p^{\obs}_{S+B}$ is Poissonian, then the domain of $S$ and $B$ would be $\mathbb{R}^n$ but the domain of the observation $X$ would be $\mathbb{N}_0^n$. However, addition/subtraction of values $B$, $S$ and $X$ must be mathematically well-defined, and the background estimator must be applicable to the domain of $B$ and $S$ as well the domain of $X$. In practice this is typically the case, though.

Under these conditions, we can now construct a forward model of the experiment without having a parameterised background model. To do this, we make the background estimator $f_{\bg}$ a (virtual) part of the experiment.\footnote{In a Bayesian context, one can also think of $f_{\bg}$ as a hyperparameter.} We replace our original observation $X$ by a virtual observation $(X', \widehat{B})$:
\begin{equation}
\begin{aligned}
    X' & \equiv X - f_{\bg}(X) \\
    \widehat{B} & \equiv f_{\bg}(X).
\end{aligned}
\end{equation}
We also define
\begin{equation}
    S'_\theta \equiv S_\theta - f_{\bg}(S_\theta).
\end{equation}
Due to the linearity (Eq.~\ref{eq:bg_est_linear}), unbiased nature (Eq.~\ref{eq:bg_est_approx}) and effectiveness (Eq.~\ref{eq:bg_est_noiseremoval}) of the background estimator we can approximate $X$ as
\begin{equation}
\begin{aligned}
    X & = X' + f_{\bg}(X) \\
      & = X' + f_{\bg}(B + N) + f_{\bg}(S) \\
      & \approx X' + B + f_{\bg}(S)
\end{aligned}
\end{equation}
and so (due to Eq.~\ref{eq:signal_decomposition}) then approximate $N$ as
\begin{equation}
\begin{aligned}
    N & = X - S - B \\
    & \approx X' + B + f_{\bg}(S) - S - B \\
    & = X' - S'_{\theta}.
\end{aligned}
\end{equation}
Now we can write an approximate but unbiased statistical model $p^{\obs}_{\theta}(X')$ for $X'$ that is independent of the unknown background $B$ and that is parameterised only by the signal parameters $\theta$:
\begin{equation}
\begin{aligned}
    p^{\obs}_{\theta}(X') & = p^{\obs}_{S_{\theta} + B}(X)  \\
    & \approx p^{\noise}_{S'_{\theta} + \widehat{B}}\bigl( X' - S'_{\theta} \bigr).
\end{aligned}
\end{equation}
So if our background estimator is accurate enough, then given an actual observation $X$ we also have a good approximation for the likelihood function of the signal parameters:
\begin{equation}
\mathcal{L}_{X'}(\theta) \approx p^{\noise}_{S'_{\theta} + \widehat{B}} \bigl( X' - S'_{\theta} \bigr).
\end{equation}
Now we can apply common statistical tools to infer the signal parameters $\theta$ based on observations $X$.

Note that in any given scenario one must carefully verify that the chosen background estimator $f_{\bg}$ is indeed unbiased (Eq.~\ref{eq:bg_est_approx}) compared to the expected signals, and that is also effective (Eq.~\ref{eq:bg_est_noiseremoval}) in suppressing the observed noise. If the signal and background fluctuate on similar length- resp. frequency-scales, it may be impossible to satisfy Eq.~\ref{eq:bg_est_approx} without violating Eq.~\ref{eq:keep_some_signal}, which would result in complete loss of sensitivity. While we assume that no parameterisation of the background is available, its general properties, like relevant length, time or frequency scales can typically be determined. So it will usually be possible to check that Eq.~\ref{eq:bg_est_approx} is fulfilled.

This also means that this method is not necessarily suitable for detecting arbitrarily small signals. It is difficult to ensure that $B - f_{\bg}(B)$ cannot partially match any allowed signal shape if the signal amplitude is allowed to approach zero - which could potentially result in falsely detecting a signal. As long as the allowed signal shapes are somewhat constrained, though, this failure should be detectable during goodness-of-fit checks.

In the following we demonstrate our approach on a specific use case, in combination with Bayesian parameter inference, but the method is valid in general under the conditions listed above. 

\vspace{-3mm}
\section{Application to an Axion Haloscope}
\label{sec:application}
\vspace{-2mm}

We use (simulated) example data of the planned axion haloscope MADMAX to show the effectiveness of our approach in a practical setting, using Savitzky-Golay filters as background estimators.

\vspace{-1mm}
\subsection{The MADMAX Experiment}
\label{Subsec:madmax} \vs{-2mm}

Axions play a crucial role in the standard Peccei-Quinn solution to the strong CP problem \cite{Weinberg:1996kr, PQMechanism, PQMechanism2, WeinbergAxion, Wilczek:1977pj}. At the same time they can also be produced non-thermally in the early universe in abundances that make them a viable dark matter candidate \cite{ABBOTT1983133,DINE1983137, PRESKILL1983127}. It is possible to detect them with earth-based experiments probing their couplings to different standard model particles, e.g.~\cite{PhysRevLett.125.011801, PhysRevD.98.082004, PhysRevLett.124.171801, PhysRevLett.129.161805, doi:10.1126/sciadv.aax4539, JEDI:2022hxa, Bloch:2022kjm, PhysRevX.13.011050}, most commonly the axion-to-photon couling $\gag$.

If they make up a sizeable fraction of a homogeneous dark matter halo \cite{HOGAN1988228, PhysRevLett.124.161103, Vaquero_2019}, axion haloscopes like ADMX \cite{ADMX:2018gho, ADMX:2019uok, ADMX:2020ote, ADMX:2021nhd, ADMX:2023rsk}, ORGAN \cite{MCALLISTER201767}, HAYSTAC \cite{HAYSTAC:2018rwy, HAYSTAC:2020kwv, HAYSTAC:2023cam} or MADMAX \cite{PhysRevLett.118.091801, Millar:2016cjp, MADMAX:2019pub, Beurthey:2020yuq} have the capability of detecting or excluding axions in certain regions of the axion mass ($\ma$), $\gag$ parameter space. MADMAX will achieve this by placing a metallic mirror and several movable dielectric disks in a dipole magnetic field. The Primakoff effect leads to the emission of radio-frequency photons at the surfaces of these disks, the energy of which depends on the axion mass. These emissions are coherent due to typically very small masses (MADMAX will be sensitive around $\sim 100\;\mu$eV) of the cold axions and correspondingly huge deBroglie wavelengths of a scale bigger than the size of the experiment. The photons can interfere constructively and be resonantly enhanced by strategically placing the disks. Through adjusting disk positions, the signal enhancement can be shifted to a different frequency, i.e. axion mass, and a big parameter space can be covered.

The observed quantity in the MADMAX experiment is the signal power in consecutive frequency bins in a given frequency range. The expected signal power (spectral power density) at a specific frequency $\omega$ adheres to the following formula:
\begin{widetext}
    \begin{equation}
        P(\omega) \drm \omega = \frac{\rhoa}{\ma^2} \gag^2 B_e^2 A \beta^2(\omega) \frac{q_e}{\hbar} \times \sqrt{\frac{2}{\pi}} \frac{v(\omega)}{\sv \vlab} \exp{\left( -\frac{v(\omega)^2 + \vlab^2}{2 \sv^2} \right)} \sinh{\left( \frac{v(\omega) \vlab}{\sv^2}\right)} \drm v(\omega)
        \label{Eq:Power}
    \end{equation}
\end{widetext}
The first part (before the $\times$) determines height and position of the signal peak. The position depends exclusively on the $\ma$, which we consider free in the frequency range after background subtraction. The height depends on multiple theoretical and experimental parameters: $\rhoa$ is the local axion density, which we fix to the canonical value of $\rhoa = 0.3$\;GeV\,cm$^{-3}$, effectively assuming homogeneous dark matter made exclusively out of axions. We also fix the experimental parameters external magnetic field $B_e = 10$\;T and disk surface area $A = 1$\;m$^2$. The power boost factor $\beta^2(\omega) = 5\times 10^4$ generally depends on frequency, for simplicity we set it constant. We expect it to vary only negligibly on the scale of the axion signal width. This leaves us with the axion-photon coupling $\gag$ as the only free parameter determining the integrated axion power. It depends solely on the anomaly ratio $\frac{E}{N}$ via $\Cag$:

\begin{equation}
    g_{a\gamma}
        =
            \frac{\alpha}{2\pi f_a}
               \mathcal{C}_{a\gamma} \equiv 
               \frac{\alpha}{2\pi f_a} \left| \frac{E}{N} - 1.92 \right|\;.
\label{Eq:AxionPhotonC}
\end{equation}

$\alpha$ is the electromagnetic fine structure constant and $f_a$ the axion decay constant, which is linearly related to the axion mass \cite{GrillidiCortona:2015jxo}. In general one should combine the prior knowledge for all parameters mentioned above, however we will only consider the most general available anomaly ratio expectation for QCD axion models for now \cite{Diehl:2023uui, plakkot2021anomaly}.

The second part of Eq.~\ref{Eq:Power} determines the shape of the signal peak. The frequency $\omega$ at which the axion can be detected depends on its total energy, so axions with different relative velocities $v(\omega)$ with respect to the laboratory can be detected at different frequencies. The relationship can be calculated by equating the photon energy with the relativistic energy of the axions yielding

\begin{equation}
        v(\omega) = \begin{cases}
                    \sqrt{1 - \left(\frac{\ma c^2}{\omega h}\right)^2} \; \; & \text{if} \; \omega h > \ma c^2\\
                    0 & \text{else}
                    \end{cases}.
\end{equation}

The dark matter velocities are assumed to follow a Maxwell-Boltzmann distribution with velocity dispersion $\sv = 218\pm6$\;km\,s$^{-1}$ \cite{bovy2012milky}. Earth is moving through the dark matter halo with a relative velocity of $\vlab = 242$\;km\,s$^{-1}$ with significant seasonal variation \cite{Knirck:2018knd}. Because we do not want to consider seasonal variations here, we move this variation into the error of the dark matter velocity distribution and model it as a Gaussian. Basically, the second part of Eq.~\ref{Eq:Power} is the probability density function of a Maxwellian velocity distribution boosted by $\vlab$. To obtain the observable integrated power in a frequency bin we have to integrate the above formula over one bin.

\begin{figure*}
    \centering
    \includegraphics[width=0.75\textwidth]{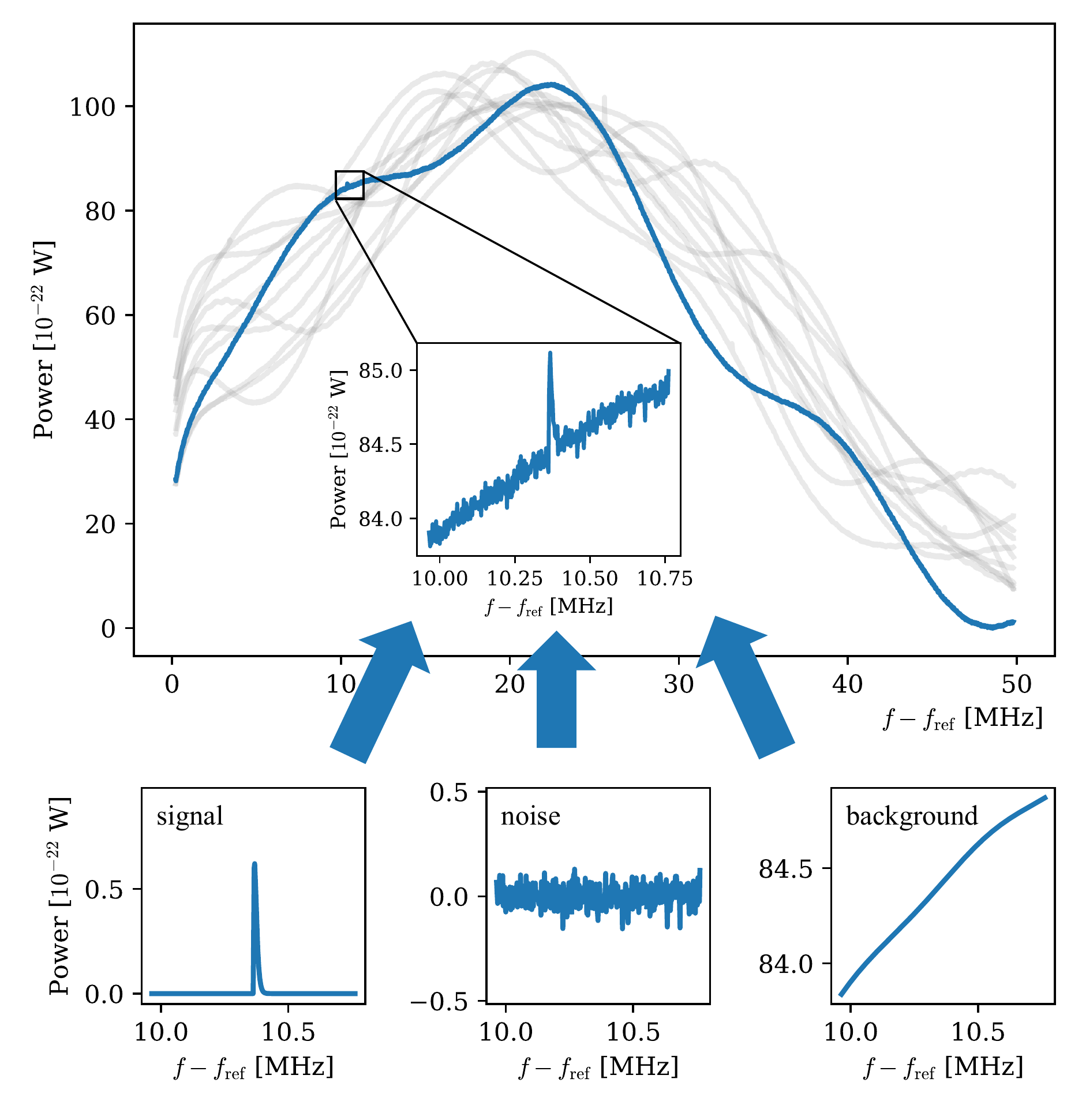}
    \vspace{-4mm}
    \caption{Components of a simulated example dataset. The upper panel shows multiple different datasets in order to visualise the variation in the simulated backgrounds, as well as a zoom into the region of the axion signal for one of them. The three lower panels contain the three components of a dataset: The simulated axion signal, uncorrelated noise and the background (from left to right).}
    \label{fig:datacomposition}
\end{figure*}

This signal sits on top of a dominant background determined by the exact characteristics of the MADMAX receiver chain. The drop off both at low and high frequencies is caused by the bandpass filter employed. The variations seen in grey in Fig.~\ref{fig:datacomposition} are exceedingly difficult to model ab-initio: Multiple components in the whole system act as correlated noise sources, the emissions of which interfere with all other sources due to the reflectivity of the system and a big coherence length at the microwave frequencies used. While all of these emissions can in theory be estimated, propagation of uncertainties in every single one of them would introduce errors in the background model of a much bigger scale than the axion signal power or the statistical noise on top of the background. 

A challenging but possible way to remove the background could be to make each measurement with the magnet turned off (no axion peak visible), the magnet turned on (axion signal visible) and subtracting the two afterwards. However, the radio-frequency interference conditions in the experiment can be slightly different if the magnet is turned on. This can potentially affect the background. Additionally even the slightest time-instability of the system between the times at which the measurements with and without magnetic field are conduced, even on levels of the noise component, would spoil this method, while it just adds a small contribution to the uncertainty when constructing a background estimator from the measurement with a magnetic field. With our approach, the only crucial requirement is that the background must not include fluctuations of the same width in frequency space as the axion signal.
\subsection{Savitzky-Golay Filters}
%
As stated above, obtaining a parametric background model for MADMAX data is likely not feasible. We thus have to rely on a non-parametric background estimator, suitable candidates being Savitzky-Golay filters.

Savitzky-Golay (SG) filters are a well-established \cite{rinnan2009review, roy2020optimal,roy2020optimal,unal2021compositional,schmid2022and}, and powerful smoothing technique often used for reducing noise in evenly spaced data while preserving the underlying smooth features \cite{savitzky1964smoothing}. The filter works by fitting low-degree polynomials to overlapping windows of data points using the method of least squares. The output of the filter is the value of the fitted polynomial at the central point of each window. The parameters of the filter are the windows length and degree of the polynomials. If these are chosen well for the given data, then SG filters can often suppress higher-frequency noise substantially while preserving the lower-frequency shape of the input. Improved versions and alternatives to SG filters have been proposed \cite{acsmeasuresciau2022}, but in our example use case we find that classical SG filters perform well.

An SG filter can be expressed compactly using matrices. Given a data vector $X$ of length $N$, we can construct a smoothed version $f_\bg(X)$ using a convolution with the filter coefficients $C$:
\begin{equation}
f_\bg(X) = X \ast C.
\end{equation}
The filter coefficients $C$ can be found by solving a linear least-squares problem. Let $A$ be a matrix of size $M \times (n+1)$, where $M$ is the (odd) window size and $n$ is the polynomial degree. Each element of $A$ is defined as:
\begin{equation}
A_{ij} = \left(i - \frac{M - 1}{2}\right)^j,
\end{equation}
where $i = 0, 1, ..., M-1$ and $j = 0, 1, ..., n$. We can find the filter coefficients $C$ by solving the following linear least-squares problem:
\begin{equation}
C = (A^T A)^{-1} A^T
\end{equation}
The first row of the resulting matrix $C$ contains the filter coefficients. Note that these coefficients are computed only once for a given window size and polynomial degree and can be used to filter the entire data vector $X$, effectively circumventing the fitting problem and resulting in high numerical performance.

SG filters are applied by convolving the filter coefficients $C$ and the input data vector $X$. They are therefore inherently linear by construction and satisfy Eq.~\ref{eq:bg_est_linear}.

By applying the SG filter to the axion haloscope data, we separate the axion signal and noise from the background, which has lower frequency characteristics. We do this without constructing a parametric model for the background. But as the signal, in contrast to the noise, is positive and not zero-symmetric, a fraction of the signal becomes part of the background estimate. This normally results in a bias that needs to be corrected.

\subsection{Data Generation and Analysis}
\label{Subsec:procedure} 
To test our bias-free signal estimation approach on the problem stated above, we simulated 1000 MADMAX-like mock-datasets. A few of these are shown as grey lines in Fig.~\ref{fig:datacomposition}. We then analysed these datasets with bias-free signal models that take the effect of the background subtraction into account and with standard signal models that do not. For our analysis we chose a Bayesian approach, however the bias-free signal estimation procedure can be applied to (and is indeed also necessary for) other inference methods, e.g., a maximum likelihood estimate of the signal parameters.

One dataset consist of 25001 data points with $2$\,kHz spacing. It has three components (see Fig.~\ref{fig:datacomposition}):
\begin{itemize}
    \item \textbf{Signal.} An axion signal following the shape of Eq.~\ref{Eq:Power}. We assume fixed $\rhoa$ and $\vlab$ and draw $\ma$, $\sv$ and $\Cag$ from the prior. For $\Cag$ we a priori exclude models with $\Cag < 0.86$ to ensure that the signals have photon couplings big enough to be detectable at the noise level given below. The $\Cag$ cutoff was chosen such that at least $99\%$ of simulated signals could be reliably recovered by the analysis. This cutoff corresponds to a SNR of $2.38$ for the signal amplitude in respect to the standard deviation of the noise. So we'll consider the scenario where an axion signal has been discovered but its quantitative properties have not been inferred yet.
    \item \textbf{Noise.}  Uncorrelated Gaussian background noise with standard deviation $\sigma = 5 \times 10^{-24}$\,W, corresponding to a realistic integration time below two weeks assuming noise temperatures below $10$\,K.
    \item \textbf{Background.}  Correlated, non-thermal background. Its shape is modelled after spectra shown in \cite{Beaujean:2017eyq}, as well as unpublished MADMAX testruns in 2022 and 2023, but does not involve fits to actual datasets. We use the following formula:
    \begin{multline}
        \mathrm{b}(f) = 10^{-20} \left[ \mathrm{erf} \left(\frac{f - f_0}{5\,\mathrm{MHz}}\right) \left(\frac{f_0}{f}\right)^3 \right. \\ 
        + \left. \exp \left( -\left(\frac{f-25\,\mathrm{MHz} (1+\frac{r_1}{15})}{20\,\mathrm{MHz} (1+\frac{r_2}{10})}\right)^2 \right) \right] \\
        + 4 \times 10^{-22} (1+r_3) \sin \left( \frac{f + r_4 f_0}{2.5\,\mathrm{MHz}} \right) \\
        + 5 \times 10^{-24} \left[ (1+r_5) \sin \left( \frac{f + r_6 f_0}{0.25\,\mathrm{MHz}} \right) \right. \\
        + \left. (1+r_7) \sin \left( \frac{f + r_8 f_0}{0.1\,\mathrm{MHz}} \right) \right],
        \label{eq:bgformula}
    \end{multline}
where $f_0 = 4.218\,\mathrm{MHz}$ and $f$ are relative frequencies and all $r_i$ are independent random variables drawn from a Gaussian $\mathcal{N}(\mu=0,\sigma=1)$. The first three lines of Eq.~\ref{eq:bgformula} determine the large-scale shape of the background, but are easy to distinguish from an axion signal with a FWHM of roughly $10$\,kHz. We therefore introduce two sine-like components with random phase, amplitudes of order of the uncorrelated noise and fixed periods of $100$\,kHz and $250$\,kHz. This choice serves as a proxy for the expected, more intricate, structure of fluctuations in physically realistic scenarios. It represents a bad-case scenario where signal extraction nevertheless remains feasible.
\end{itemize}

\begin{table}
    \centering
    \caption{Overview over the fixed parameters in the analysis and the priors used for the three non-fixed parameters. $f_{\rm min / max}$ are the extremal absolute frequencies at which the measurement
is sensitive.}
    \begin{tabular}{l l}
        \toprule
         Parameter & Value/ Prior \\ \midrule
         $\ma$ & Uniform$(f_{\rm min}, f_{\rm max})$ \\
         $\Cag$ & From \cite{Diehl:2023uui, plakkot2021anomaly}, excluding $\Cag < 0.86$ \\
         $\sv$ & Normal$(218,39)$ [km\;s$^{-1}$] \\ \midrule
         $\rhoa$ & $0.3\,$GeV$\,$cm$^{-3}$ \\
         $\beta^2$ & $5 \times 10^4$ \\
         $v_\mathrm{lab}$ & $242\,$km$\,$s$^{-1}$ \\
         $B_e$ & $10\,$T \\
         $A$ & $1\,$m$^2$ \\\bottomrule
    \end{tabular}
    \label{tab:priors}
\end{table}
\begin{figure}
    \centering
    \includegraphics[width=0.45\textwidth]{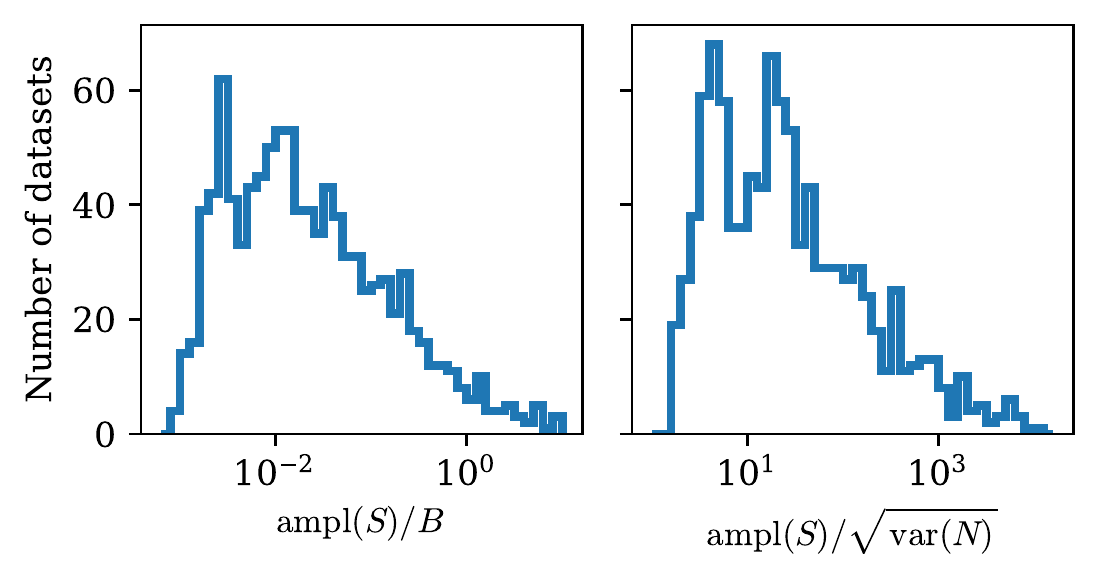}
    \vspace{-4mm}
    \caption{Distribution of the signal amplitude across our mock datasets, relative to the background and the noise level. \textbf{Left:} Signal to background ratio, given by maximal amplitude of the signal divided by the background at that frequency. \textbf{Right:} Signal to noise ratio, given by maximal amplitude of the signal divided by standard deviation of the noise component. The bimodal shape of the SNR histogram is caused by the $\Cag$ prior the signals are sampled from.}
    \label{fig:SBR+SNR}
\end{figure}

Tab.~\ref{tab:priors} summarises the relevant parameters for our analysis.

The signals in our example application are small in the sense that the median of their signal to background ratios (SBR) is at $0.016$. Fig.~\ref{fig:SBR+SNR} shows the SBRs and signal to noise ratios (SNRs) for our mock datasets, represented by signal amplitude divided by background at the signal frequency and signal amplitude divided by noise standard deviation respectively.

We filter the data consisting of these three components with a fourth-order SG filter of a width of $221$ data points, cutting away the first and last 110 \mbox{data points} due to boundary effects of the filter. Subtracting the filtered from the raw data removes the third, correlated background component almost completely, but also slightly distorts the signal shape, as shown below. The parameters of the SG fit were chosen to yield a good reduction of the background while leaving signal and noise almost unchanged. The exact choice of SG parameters does not affect the validity of our approach. On the contrary: While the parameter bias induced by previous approaches depends on filter parameters, our bias-free inference is independent of them, as long as the SG filter is compatible with the background (see Sec.~\ref{sec:approach}).

\begin{figure*}
    \centering
    \includegraphics[width=0.99\textwidth]{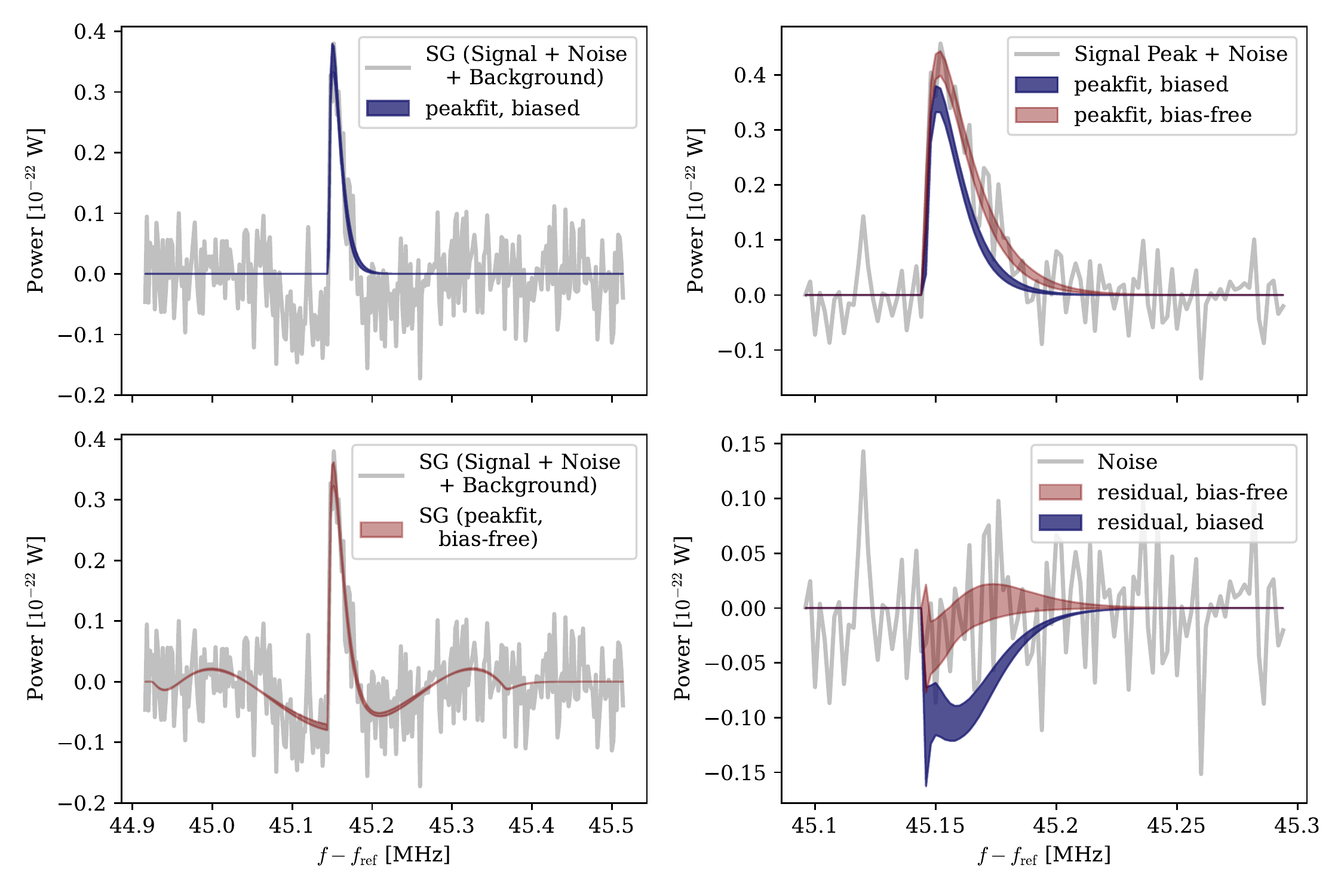}
    \vspace{-4mm}
    \caption{Effect of using a bias-free signal model on the fit on the data. $68\%$ central posterior predictive intervals for the bias-free and biased fit are plotted in red and blue respectively. \textbf{Top Left:} Data after background reduction via SG filter and biased fit without SG filter applied. \textbf{Bottom Left:} Same data and bias-free fit with SG filter applied. \textbf{Top Right:} Signal and noise components of the data as well as biased and bias-free peak-fit without applying an SG filter. \textbf{Bottom Right:} Deviation of posterior predictive of bias-free and biased peak-fit from the true signal ("residual") compared to noise component of the data. The peak-fits in all panels have been performed on background reduced data as shown in the left panels. The data components shown in the right panels are for comparison only.}
    \label{fig:correction}
\end{figure*}

We use Eq.~\ref{Eq:Power} integrated over the frequency bins as the signal model for our Gaussian likelihood. As standard deviation we take the amplitude of the uncorrelated noise, which is unknown in realistic scenarios and therefore has to be inferred from the data. If we would simply take the standard deviation of the background reduced data, the presence of a signal would lead to a slight bias for the noise towards higher values. To prevent the signal from biasing our noise-level estimation we first use the SG filter to remove a background estimate from the data. Then we partition each spectrum into three frequency regions of equal size (roughly $8333$~bins). Because the width of each of these regions is much bigger than that of the localised filtered signal, it can only be present in up to two of the regions simultaneously. We select the region with the smallest standard deviation for our noise-level estimation. This removes the bias caused by the presence of the signal. In our test case the inferred noise level has a negligible difference compared to the ground truth. This procedure works only if the background filter is unbiased and the noise level of the residuals is constant, which we verified for our test case. In some applications relative residuals might need to be used, especially for frequency dependent $N$ or if background oscillations depend on its absolute magnitude.

For Monte Carlo analysis we use the Reactive Nested Sampling algorithm \cite{NestedSampling1, NestedSampling2, NestedSampling3} via the Bayesian Analysis Toolkit in Julia (BAT.jl) \cite{Schulz:2021BAT}.

\subsection{Requirement Validation}

Sec.~\ref{sec:approach} states several requirements for our method to be applicable. This subsection will show that they are all fulfilled in our example use case.

The first two requirements are fulfilled by construction of our mock datasets. The red curve in Fig.~\ref{fig:correction} (bottom left panel) corresponds to $S - f_\bg(S)$ and clearly shows, that it is not equal to zero everywhere. We also already presented a parameterisation of $S$ in Sec.~\ref{Subsec:madmax}. It remains to be shown that Eqs.~\ref{eq:bg_est_approx} and \ref{eq:bg_est_noiseremoval} approximately hold, which hinges on the choice of SG filter parameters.

\begin{figure}
    \centering
    \includegraphics[width=0.49\textwidth]{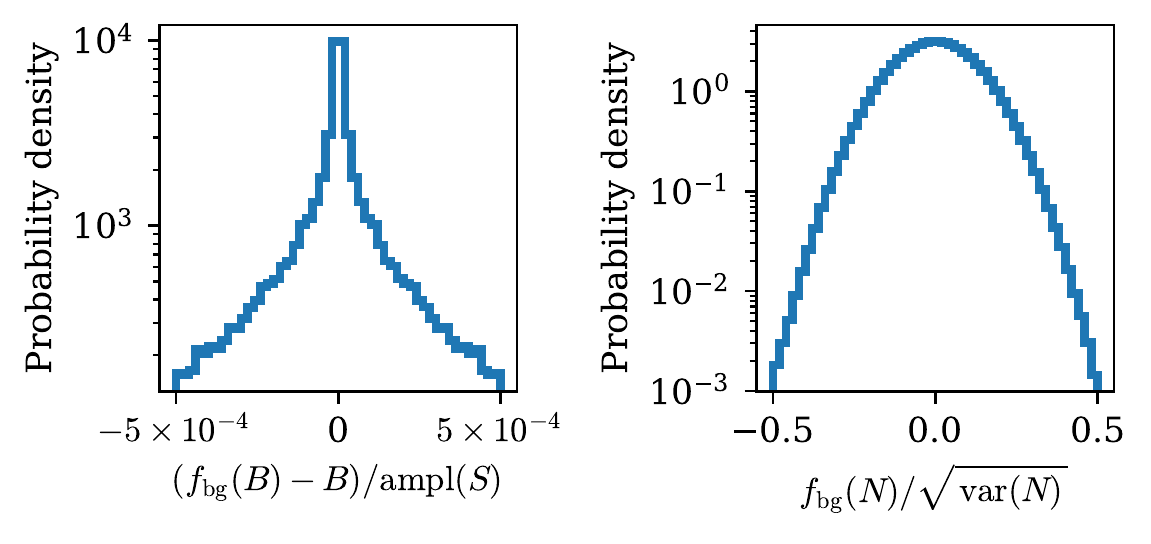}
    \vspace{-8mm}
    \caption{
    Validation of the chosen SG filter parameters in terms of background estimation and noise removal. \textbf{Left:} Deviation between filtered and unfiltered true backgrounds for all data points normalised to the signal amplitude for each mock dataset (see Eq.~\ref{eq:bg_est_approx}). \textbf{Right:} Comparison between the filtered noise for all data points and the standard deviation of the unfiltered noise component for each mock dataset (see Eq.~\ref{eq:bg_est_noiseremoval}).}
    \label{fig:SG_bias}
\end{figure}

We consider the SG filter to be a good background estimator if $|B - f_{\bg}(B)| \ll |\mathrm{ampl}(S)|$ and to be efficient at removing the noise component if $|f_{\bg}(N)| \ll \sqrt{\mathrm{var}(N)}$. Fig.~\ref{fig:SG_bias} shows the distribution for both of these conditions using an SG filter with a width of $221$ data points and polynomial order $4$ on our mock datasets.

Fig.~\ref{fig:SG_bias}, left panel, demonstrates the chosen SG filter to be an excellent background estimator with practically all residuals being below $5\times 10^{-4}$ of the signal amplitude.

The SG filter applied to noise retains the Gaussian behaviour of our noise component, but reduces its standard deviation to $ 0.13 \sqrt{\mathrm{var}(N)}$, so by almost one order of magnitude, as can be seen in Fig.~\ref{fig:SG_bias}, right panel.

Our chosen SG parameters therefore represent a filter that fulfils all requirements listed in Sec.~\ref{sec:approach}.

\subsection{Results}
\label{Subsec:results} 

Fig.~\ref{fig:correction} shows the result for one of the datasets. We demonstrate the effect of the bias-free vs biased method qualitatively using $68$ percentile central posterior intervals. When the standard signal model is used to fit the signal peak (Fig.~\ref{fig:correction}, top left), the effect of the background removal via SG filter cannot be modelled. Due to the presence of the signal, the background around the signal is overestimated, leading to a systematic decrease in signal height and adjacent data points below the baseline. A peak-fit is perfectly capable of fitting this modified signal, but cannot fit the surrounding data points - and therefore does not retain the true signal parameters. We will see this in the following.

The bottom-left plot in Fig.~\ref{fig:correction} shows the same, but with a bias-free fit on the signal peak that takes the effect of SG filtering into account. We obtain a good fit over the whole frequency range, which displays the characteristic effect of an SG filter on the signal. Unbiased signal parameters can be obtained based on this fit, as Fig.~\ref{fig:correction}, top right shows. The actual, non-filtered and background free signal peak is fitted well by the posterior predictive central interval of the bias-free peak-fit on which no SG filter has been applied. The standard, biased peak-fit however underestimates signal height and width. This underestimation is made more visible in Fig.~\ref{fig:correction}, bottom right, where the deviation relative to the true signal is shown in comparison with the noise level, both for the biased and the bias-free peak-fit. While the $68$ percentile of the bias-free fit includes the true signal over almost the full frequency range, the standard, biased fit displays significant deviation.

\begin{figure}
    \centering
    \includegraphics[width=0.45\textwidth]{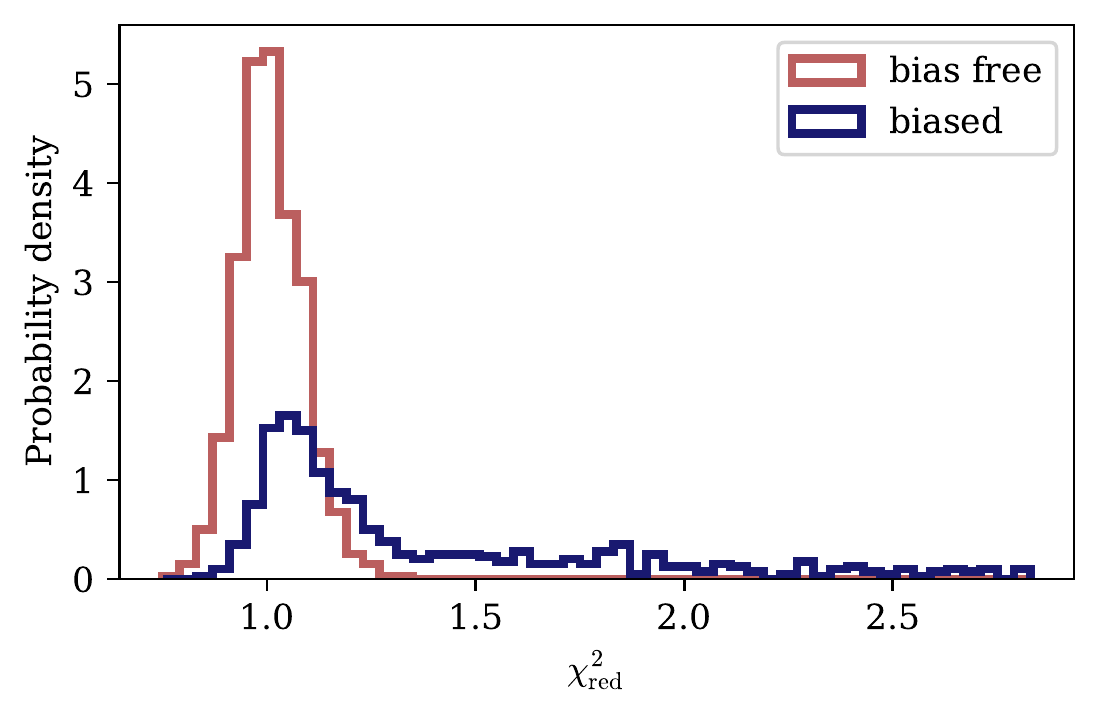}
    \vspace{-4mm}
    \caption{Reduced chi-square distributions for all mock datasets, obtained from fits using the standard biased method (blue) as well as our bias-free method (red). The reduced chi-square was calculated for the posterior median using the inferred noise level as described above and only in a range of 350 data points around the known signal frequency, because only in this region are pronounced differences between the biased an our bias free method to be expected.}
    \label{fig:chi2}
\end{figure}

As a first quality-of-fit check, we show a histogram of reduced chi-square values for all mock datasets in Fig.~\ref{fig:chi2}. The distribution of chi-square values for our bias-free method is compatible with unity ($1.005 \pm 0.08$). For the biased method, however, we observe a distribution of chi-square values systematically above one. While most datasets display reasonably good $\chi^2_\mathrm{red} \in [0.8, 1.5]$, some reduced chi-square values exceed even the upper bound of Fig.~\ref{fig:chi2}, indicating extremely bad fits.

\begin{figure}
    \centering
    \includegraphics[width=0.45\textwidth]{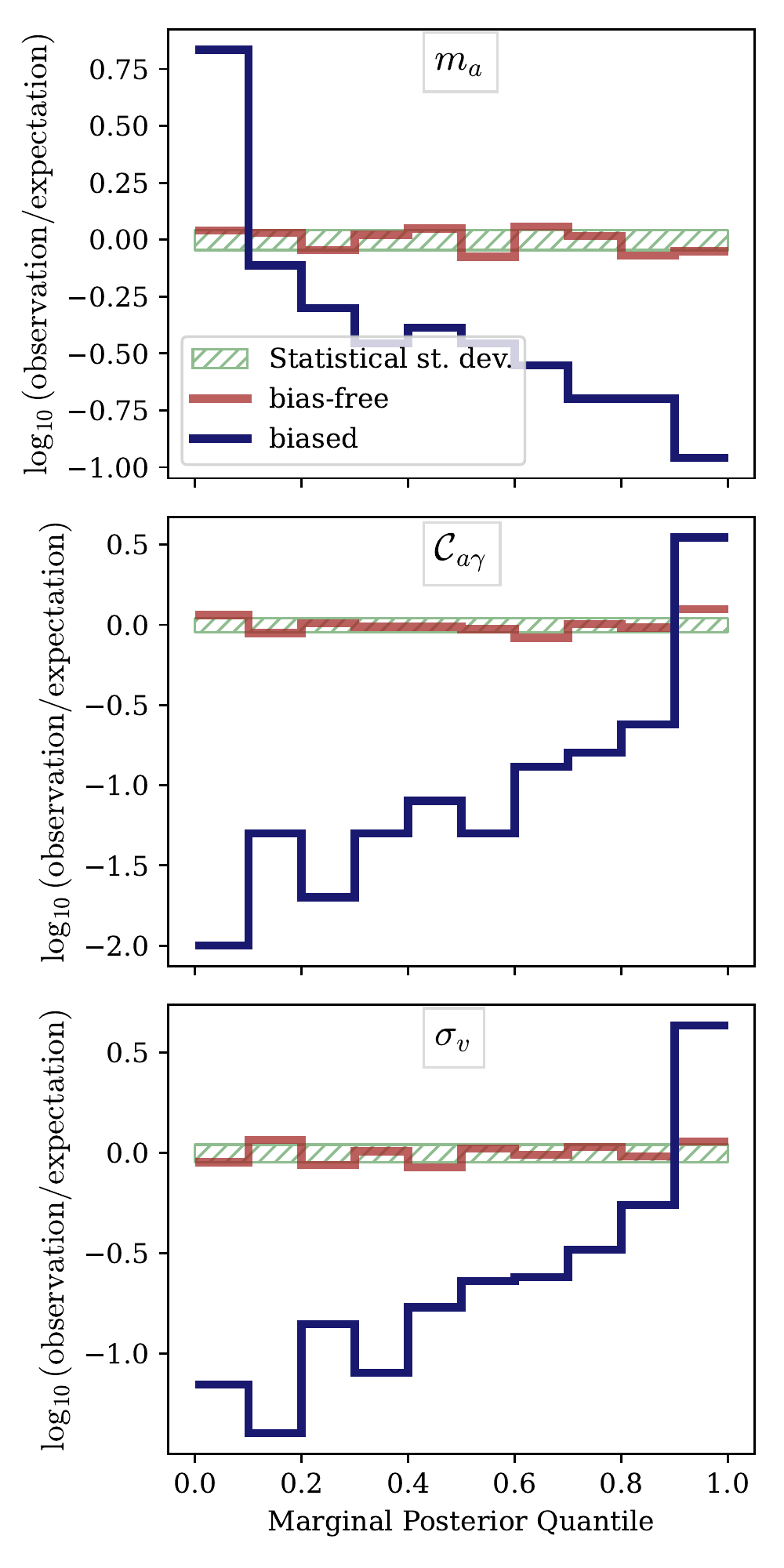}
    \vspace{-4mm}
    \caption{Distribution of the posterior quantile bin in which the true signal parameter resides, for the three parameters. The red curves show the bias-free parameter inference, the blue curves show the biased one. For 1000 mock-datasets and 10 quantile bins $100 \pm 10$ cases are expected for each bin, which is shown as the green dashed region. The width of the green band corresponds to the expected statistical fluctuation. The distribution of the parameters in our bias-free approach shows no significant deviations from the expectation. For the standard, biased approach, we observe a significant and systematic shift away from the true signal parameters. In this case almost all true signal parameters reside in the first or last quantile of the posterior distribution.}
    \label{fig:diagnostic}
\end{figure}

\begin{table}
    \centering
    \caption{Coverage properties of the Bayesian credible intervals: The table shows the percentage of cases in which true signal parameter resides in the $68\%$ (resp. $95\%$ and $99.7\%$) posterior credible interval, for each of the three free fit parameters.}
    \begin{tabular}{lcccccc}
        \toprule
        \multirow{2}{*}{} & \multicolumn{3}{c}{Coverage biased} & \multicolumn{3}{c}{Coverage bias-free} \\
          & $68\%$ & $95\%$ & $99.7\%$ & $68\%$ & $95\%$ & $99.7\%$  \\ \midrule
         $m_a$      & 0.241 & 0.405 & 0.589 & 0.690 & 0.945 & 0.994  \\
         $\Cag$      & 0.055 & 0.131 & 0.426 & 0.643 & 0.925 & 0.993  \\
         $\sigma_v$ & 0.141 & 0.25 & 0.387 & 0.669 & 0.946 & 0.992  \\ \bottomrule
    \end{tabular}
    \label{tab:coverage}
\end{table}

To further asses the quality of our fits, we perform Bayesian coverage testing and show that the Bayesian posterior are compatible with the true parameter values, but only when using our bias-free approach. From a frequentist perspective, repeating an unbiased analysis multiple times should lead to the true signal parameters being uniformly distributed over all marginal posterior quantiles. As we do have access to the ground truth of the signal parameters (using simulated data), and as we have 1000 equivalent mock-datasets at our disposal, we can verify our Bayesian results in this manner. The outcome is shown in Fig.~\ref{fig:diagnostic}. The standard, biased peak-fit that does not take the effect of the SG filter into account systematically underestimates $\Cag$ and $\sigma_v$. It also shifts the axion mass to slightly larger values. The bias-free fit, in comparison, shows no significant deviation from the expected uniform distribution. A more standard test to quantify coverage is given in Tab.~\ref{tab:coverage}. With the standard biased method, only a small fraction of true signal parameters lies within the $68\%$, $95\%$ or $99.7\%$ credible intervals of the posterior, whereas the coverage for our bias-free method is close to the expected values.

\begin{figure}
    \centering
    \includegraphics[width=0.45\textwidth]{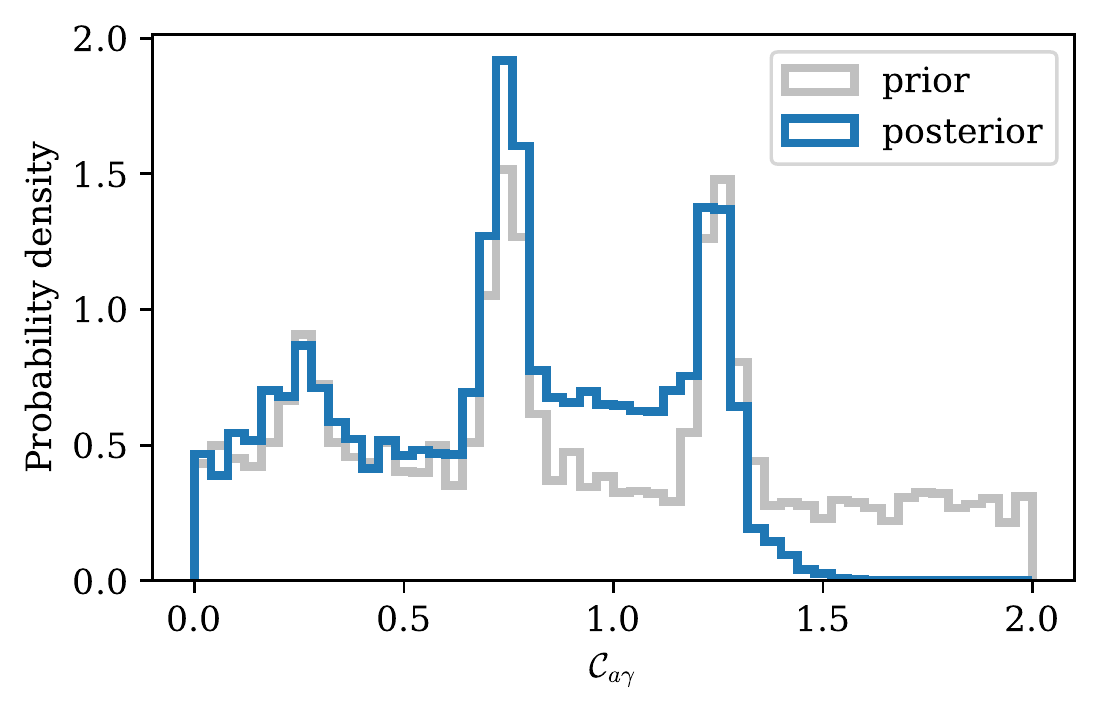}
    \vspace{-4mm}
    \caption{Posterior and prior probability density for signal parameter $\Cag$ for a mock dataset with no simulated axion present. The prior extends far beyond the region shown for $\Cag > 2$. As expected, the posterior is compatible with a no-signal hypothesis, sets an upper limit on $\Cag$, and shows no gain of information in respect to the prior at low values.}
    \label{fig:no_signal_posterior}
\end{figure}

Fig.~\ref{fig:no_signal_posterior} shows the prior and posterior distributions for a simulated dataset with no axion present. The probability density of the posterior is clearly severely suppressed for $\Cag > 1.5$, which hints at the limit setting potential of a MADMAX-like experiment with parameters as chosen above. Below $\Cag < 1.4$ posterior and prior follow similar distributions, no information has been gained. The $\Cag$ posterior is therefore compatible with a no-axion scenario, as expected.  

Positive statistical fluctuations in several adjacent bins can potentially mimic faint signals in scenarios with no detectable axion present. To avoid erroneous detections in these scenarios it is crucial to not underestimate the noise level implemented in the likelihood and to design a statistically sound detection criterion. The latter is a highly relevant question on its own right, but requires extensive validation and is beyond the scope of the present paper. For this reason we a priori assume $\Cag > 0.86$ in our mock dataset, corresponding to detectable signal strengths in the present analysis (compare Tab.~\ref{tab:priors}). One could, however, reasonably assume that the inclusion of the presented approach could improve the sensitivity of detection procedure, as the modified signal model will more closely resemble the respective features in the data. We leave the exploration of this for future work.

\section{Summary}
\label{sec:Summary_and_Outlook}

We present a method to fit small-amplitude signals on top of an unparameterised background in an unbiased fashion. The method is based on fairly weak assumptions about the problem, making it applicable in a wide variety of scenarios: we mainly require the existence of an unbiased and linear background estimator and an a-priori parameterisation of the signal. This enables us to virtually incorporate the background estimator into the measurement process and reduces the problem to modelling a background-free signal in the presence of symmetric noise, i.e., noise that has an expectation value of zero.

To evaluate the method in a practical real-world application, we applied it to signal estimation on simulated data for the planned MADMAX axion haloscope. The MADMAX experiment aims to detect a small, peaked axion dark matter signal in the presence of a challenging radio-frequency background that is difficult to model from first principles, so it is a suitable candidate for the presented approach. For 1000 MADMAX-like mock datasets we used Savitzky-Golay filters as background estimators and inferred the signal parameter in a Bayesian fashion using nested sampling. The results verify the bias-free approach presented here empirically and show that the signal parameter estimates are indeed unbiased. The true signal parameters are within the 68$\%$ central posterior predictive limits almost everywhere and the deviations of inferred parameters from the ground truth fall within the range of expected statistical fluctuations.

For comparison, we performed the signal parameter estimation in the standard, biased fashion. Here the results do indeed show significant systematic deviations of the inferred signal parameters from the ground truth. The bias free method is thus both effective and necessary.

\section*{Acknowledgements}

The authors would like to thank Allen Caldwell and Frank Steffen for helpful comments on the manuscript. Jakob Knollmüller acknowledges funding by the Deutsche Forschungsgemeinschaft (DFG, German Research Foundation) under Germany´s Excellence Strategy – EXC 2094 – 390783311.
Jakob Knollmüller also acknowledges funding by the European Research Council (ERC) under the European Union’s Horizon 2020 research and innovation programme (Grant agreement No. 101071643).


\bibliographystyle{utphys86} 
\bibliography{main}








%
%





\end{document}

%% file: header.tex
\usepackage{amsmath}	
\usepackage{amsthm}		
\usepackage{amssymb}	
\usepackage{datetime}
\usepackage{graphicx}
\usepackage{color}
\usepackage{verbatim}
\usepackage{bm}

\usepackage[colorlinks=true, citecolor=midblue, linkcolor=midblue, urlcolor=midblue]{hyperref}

\usepackage{setspace}
\usepackage[flushleft]{threeparttable}
\usepackage{multirow}
\usepackage{tabularx, booktabs}

\usepackage[verbose]{placeins}
\usepackage{widetext}
\usepackage[numbers]{natbib}

\definecolor{grey}{rgb}{0.4,0.4,0.4}
\definecolor{dullmagenta}{rgb}{0.4,0,0.4}
\definecolor{darkblue}{rgb}{0,0,0.4}
\definecolor{midblue}{rgb}{0,0,0.5}
\definecolor{midred}{rgb}{0.5,0,0}
\definecolor{orange}{rgb}{1,0.5,0}
\definecolor{lightbrown}{rgb}{0.75,0.5,0.25}
\definecolor{tan}{cmyk}{0.14,0.42,0.56,0}
\definecolor{djunglegreen}{cmyk}{0.99,0,0.52,0}
\definecolor{lightgreen}{rgb}{0,1,0}
\definecolor{olivegreen}{cmyk}{0.64,0,0.95,0.40}
\definecolor{midgreen}{rgb}{0.0,0.675,0.0}
\definecolor{darkgreen}{rgb}{0,0.5,0}

\newcommand{\vs}{\vspace}

\newcommand{\drm}{\ensuremath{\mathrm{d}}}

\smallskipamount

\settimeformat{ampmtime}

\usepackage{float}



%% file: mycommands.tex
\newcommand{\gag}{g_{a\gamma}}
\newcommand{\rhoa}{\rho_a}
\newcommand{\ma}{m_a}
\newcommand{\sv}{\sigma_v}
\newcommand{\vlab}{v_{\rm lab}}
\newcommand{\Cag}{\mathcal{C}_{a \gamma}}

\newcolumntype{Z}{>{\raggedleft\arraybackslash}X}

\newcommand{\obs}{\mathrm{obs}}
\newcommand{\noise}{\mathrm{noise}}
\newcommand{\bg}{\mathrm{bg}}